\documentclass[10pt]{article}

\usepackage{amsmath}
\usepackage{amssymb}

\usepackage{graphicx}

\usepackage[numbers,square]{natbib}
\begin{document}

\begin{flushleft}
{\Large
\textbf{Microbiome profiling by Illumina sequencing of combinatorial sequence-tagged PCR products}
}
\\
Gregory B. Gloor\,$^{1,\ast}$,
Ruben Hummelen\,$^{2,3}$,
Jean M. Macklaim\,$^{1,2}$,
Russell J. Dickson\,$^{1}$
Andrew D. Fernandes\,$^{1,4}$,
Roderick MacPhee\,$^{2,5}$,
Gregor Reid\,$^{1,2,5,6}$
\\
\bf{1} Department of Biochemistry, University of Western Ontario, London, ON, Canada
\\
\bf{2} Canadian Research \& Development Centre for Probiotics, Lawson Health Research Institute, London, Canada
\\
\bf{3} Department of Public Health, Erasmus MC, University Medical Centre Rotterdam, The Netherlands
\\
\bf{4} Department Applied Mathematics
\\
\bf{5} Department of Microbiology \& Immunology, University of Western Ontario, London, ON, Canada
\\
\bf{6} Department of Surgery, University of Western Ontario, London, ON, Canada
\\
$\ast$ E-mail: ggloor@uwo.ca
\end{flushleft}

\section*{Abstract}
We developed a low-cost, high-throughput microbiome profiling method that uses combinatorial sequence tags attached to PCR primers that amplify the rRNA V6 region. Amplified PCR products are sequenced using an Illumina paired-end protocol to generate millions of overlapping reads. Combinatorial sequence tagging can be used  to examine hundreds of samples with far fewer primers than is required when sequence tags are incorporated at only a single end. The number of reads generated permitted saturating or near-saturating analysis of samples of the vaginal microbiome. The large number of reads allowed an in-depth analysis of errors, and we found that PCR-induced errors composed the vast majority of non-organism derived species variants, an observation that has significant implications for sequence clustering of similar high-throughput data. We show that the short reads are sufficient to assign organisms to the genus or species level in most cases. We suggest that this method will be useful for the deep sequencing of any short nucleotide region that is taxonomically informative; these include the V3, V5 regions of the bacterial 16S rRNA genes and the eukaryotic V9 region that is gaining popularity for sampling protist diversity.

\section*{Introduction}
Microbiome profiling is used to identify and enumerate the organisms  in samples from diverse sources such as soil,  clinical samples and oceanic environments \cite{Andersson:2008,Lauber:2009,Polymenakou:2009}. This profiling is an important first step in determining the important bacterial and protist organisms in a biome and how they interact with and influence their environment. 

Microbiome profiling is usually achieved by sequencing PCR-amplified variable regions of the bacterial 16S and of the protistan small subunit ribosomal RNA genes \cite{Hamady:2009, Amaral-Zettler:2009}. Other sequences, such as the GroEL genes may also be targeted for independent validation \cite{Schellenberg:2009}. The microbial profile of a sample may be determined by traditional Sanger sequencing, by terminal restriction length polymorphism analysis or by denaturing gradient gel electrophoresis (reviewed in \cite{Nocker:2007}. The recent introduction of massively parallel 454 pyrosequencing  has resulted in a radical increase in the popularity of microbiome profiling because a large number of PCR amplicons can be sequenced at for a few cents per read  \cite{Petrosino:2009, Hamady:2009}.  However, while constituting a tremendous improvement over previous methods, pyrosequencing is constrained by cost limitations and a relatively high per-read error rate. The high error rate has led to some discussion in the literature about the existence and importance of the `rare microbiome' \cite{Reeder:2009}. New methods for analyzing pyrosequencing output suggest that much of the rare microbiome is composed of errors introduced by the sequencing method \cite{Quince:2009}.

Until recently, the Illumimna sequencing-by-synthesis method of parallel DNA sequencing was thought to be unsuitable for microbiome profiling because the sequencing reads were too short to traverse any of the 16S rRNA variable regions. This can be partially circumvented by identifying maximally informative sites for specific groups of organisms (eg. \cite{Pawlowski:2010}). A recent report demonstrated that short sequences derived from Illumina sequences could be used for robust reconstruction of bacterial communities. This group used Illumina  sequencing to determine the partial paired-end sequence of the V4 16S rRNA region in a variety of samples using single-end sequence tagged PCR primers \cite{Caporaso:2010}.  

Here we report the methods used to perform microbiome analysis of the V6 region of 272 clinical samples using the Illumina sequencing technology. We used paired-end sequencing in combination with unique sequence tags at the 3` end of each primer. The overlapping paired-end reads gave us complete coverage of the V6 region. The combination of sequence tags at each end of the overlapped reads allowed us to use a small number of primers to uniquely tag a large number of samples. The Illumina sequencing method generated $\sim 12$ million useable reads at a cost of $\sim 0.03$ cents per read, an approximate order of magnitude cheaper than the per-read cost of pyrosequencing. The cheaper per-read costs allows economical experiments on large numbers of samples at very large sequencing depths. Since Illumina sequencing is now capable of $\sim100$ nt long reads from each end of a DNA fragment, the methods described here can be easily adapted for paired-end sequencing of the microbial V3, V5, V6 and the eukaryotic V9 rRNA regions. Similarly to others \cite{Caporaso:2010}, we found that methods used to analyze pyrosequencing microbiome data were often unsuitable for reads generated by Illumina sequencing and we present a workflow that can be used for rapid and robust generation of the relative abundance of organisms in each sample. 

Importantly, we found that the Illumina sequencing method has an exceedingly low error rate and that the majority of errors arise during the PCR amplification step. We argue that the error profile has profound implications for choosing the appropriate seed sequence for clustering using the data generated by Illumina sequencing. 

\section*{Results}
\subsection*{Description of the data}

The DNA samples analyzed by this method were derived from a study designed to examine the vaginal microbiota in HIV+ women in an African population. A separate manuscript details the clinical findings of the study \cite{Hummelen:2010}. In all we analyzed 272 clinical samples by a single Illumina paired-end sequencing run.

\begin{figure}[ht]
\begin{center}
\includegraphics[width=0.7\textwidth]{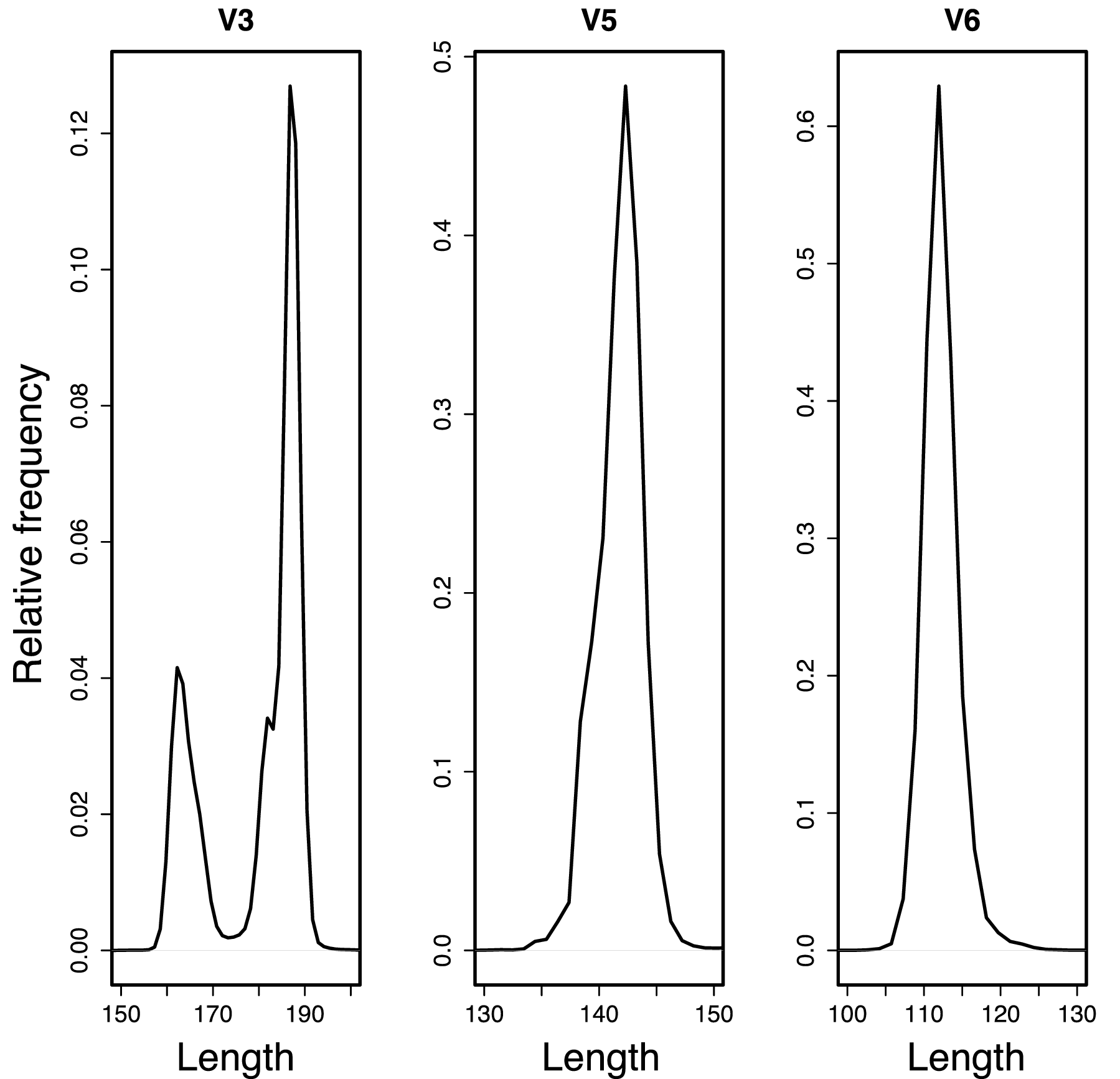}
\end{center}
\caption{{\bf Expected amplified product size using constant regions flanking eubacterial variable regions.} }
\label{vsize}
\end{figure}

The Illumina sequencing platform is currently restricted to read lengths of $\sim 100$ nucleotides from each end of a DNA fragment, and was limited to $\sim$75 nt at the time of experimental design. Thus, a paired-end sequencing run could only traverse the short 16S variable regions: V3, V5 and V6. The expected distribution of amplified fragment sizes, including the primer,  for each variable  region is shown in Figure \ref{vsize}. We decided to use the V6 region for two main reasons. First, the V6 region  was expected to produce amplified fragments  between 110 and 130 bp, ensuring that  the majority of paired-end reads would overlap. Secondly,   the V6 region provided resolution for a number of organisms of interest in our samples down to the species and in some cases the strain level \cite{Huse:2008}.  The Illumina platform currently provides reads long enough to overlap in either the bacterial V3 \cite{Huse:2008, Andersson:2008} or V5 regions or in  the eukaryotic V9 region \cite{Amaral-Zettler:2009}. We suggest the region(s) chosen for sequencing should be characterized for the resolution of taxa of interest, and several studies have examined this in detail \cite{Pawlowski:2010, Nocker:2007}.

\subsection*{PCR Primer characterization}
The primers were located within two conserved 16S rRNA segments that flanked the V6 region. The left and right primer sequences mapped to the 967-985 (CAACGCGARGAACCTTACC) and 1078-1061 (ACAACACGAGCTGACGAC) using the coordinates on the \emph{Escherichia coli} 16S rRNA segment, and were chosen to amplify the majority of species expected to be found in the vaginal environment. The potential ability to amplify the regions flanking the V6 region was tested computationally by two methods. First, the primers were assessed using the probematch service from the Ribosomal Database Project \cite{Cole:2009}. The forward and reverse primers were found to match 96.8\% and 99.3\% of the good quality, long 16S rRNA sequences with 2 or fewer mismatches. The forward primer was strongly biased against amplification of sequences in the Tenericutes and Thermotogae phyla, amplifying 201/1438 and 8/82 in these groups. The reverse primer was unbiased. Secondly, we used a  method similar to Wang and Qian  \cite{Wang:2009}. Unaligned 16S rRNA sequences were downloaded from the Ribosomal RNA Database Project \cite{Cole:2009} and the 187260 sequences longer than 1400 nt were extracted. Sequences of this length are nearly-full length and are expected to contain the V6 region. These sequences were filtered to remove entries where the only entry on the annotation line was `unidentified bacterium' or `uncultured bacterium', leaving 97987 entries. Approximate string matching (agrep) with the TRE regular expression library \cite{tre:2010} was used to determine that the left and right primers matched 94101 and 96432 of 97987 sequences with the requirements of perfect matching at the 5 nucleotides at the 3` end and up to 2 mismatches in the remainder of the primer.  Using this measure, the left primer matched over 96\% and the right primer over 98\% of the sequences in the dataset. However, as shown in Table \ref{hits}, either the left or right the primers did not match the majority of sequences annotated as \emph{Sneathia}, \emph{Leptotrichia}, \emph{Ureaplasma} or \emph{Mycoplasma}. We found that relaxing the parameters somewhat resulted in matching to the majority of species in these groups (Table \ref{hits}). We suggest that these primers would allow amplification of the majority of species in each of these groups, but that amplification may occur at lower efficiencies in some groups.
\begin{table}[ht]
\caption{\bf{Number of species matching each primer in a filtered RDP dataset}}
\begin{tabular}{lccc}
Taxon & Total species & Left$^1$ & Right$^a$\\ \hline 
\emph{Escherichia} & 322 & 320 & 318 \\
\emph{Citrobacter} & 113 & 111 & 110 \\
\emph{Bacteroides} & 275 & 265 & 270 \\
\emph{Streptococcus} & 1249 & 1243 & 1244 \\
\emph{Staphylococcus} & 704 & 696 & 694\\
\emph{Lactobacillus} & 1922 & 1908 & 1910 \\
\emph{Lachnospiraceae} & 82 & 82 & 82 \\
\emph{Peptostreptococcus} & 28 & 28 & 28 \\
\emph{Anaerococcus} & 29 & 29 & 29 \\
\emph{Megasphaera} & 38 & 38 & 38 \\
\emph{Dialister} & 21 & 21 & 21 \\
\emph{Candidatus} & 579 & 377 & 566 \\
\emph{Mobiluncus} & 25 & 25 & 25 \\
\emph{Propionibacteriaceae} & 12 & 12 & 12 \\
\emph{Bifidobacterium} & 146 & 145 & 143 \\
\emph{Porphyromonas} & 111 & 109 & 111 \\
\emph{Prevotella} & 269 & 264 & 264 \\ 
\emph{Fusobacterium} & 103 & 102 & 103 \\ 
\emph{Sneathia} & 4 & 4 & 0(4)$^b$ \\
\emph{Leptotrichia} & 60 & 60 & 1(58)$^b$  \\ 
\emph{Gardnerella} & 3 & 3 & 3 \\
\emph{Ureaplasma} & 36 & 0(34)$^c$ & 36 \\
\emph{Mycoplasma} & 414 & 95(331)$^d$ & 336 \\
\end{tabular}
\begin{flushleft}
$^a$ \footnotesize{number of hits with identity at the 3' 5 nucleotides and up to 2 mismatches in the rest of the primer:}\\
$^b$ \footnotesize{number of hits requiring identity at the 3' 4 nucleotides:}\\
$^c$ \footnotesize{number of hits allowing 3 mismatches and identity at the 3' 5 nucleotides:}\\
$^d$ \footnotesize{number of hits allowing 4 mismatches and identity at the 3' 5 nucleotides.}
\end{flushleft}
\label{hits}

\end{table}%

The primers were tested for their ability to amplify the 16S rRNA V6 region of \emph{Lactobacillus iners}, \emph{Lactobacillus rhamnosus}, \emph{Gardnerella vaginalis} and \emph{Atopobium vaginae}. All were amplified equivalently using the following PCR parameters: denaturation $94^{\circ}$, annealing $51^{\circ}$, extension $72^{\circ}$ all for 45 seconds over 25 amplification cycles.

\subsubsection*{Sequence tag choice} The Illumina sequencing platform uses dye-terminated primer extension to sequence DNA \cite{Bentley:2008} and the base-calling algorithm uses the intensities from the first several nucleotides incorporated to normalize the fluorescent signal from subsequent nucleotide incorporation events \cite{Smith:2008,Whiteford:2009}. Thus, we chose sequence tags to ensure all 4 nucleotides were represented in each of the first four positions of the primers using parameters similar to those in the barcrawl program \cite{Frank:2009}. This was achieved, in part, by varying the length of the tags between 3 and 6 nucleotides long. The tag length variation was expected to reduce the likelihood that adjacent spots on the Illumina solid support would be scored as one during the sequencing of the amplification primers \cite{Smith:2008,Whiteford:2009}. All sequence tags were checked with a primer design program to ensure that they would not induce primer-dimer formation \cite{Engels:1993}. The sequence tags are given in Table \ref{barcode}. The right-side sequence tags can be uniquely identified if they are full-length, or if they are truncated by 1 nucleotide, as commonly occurs during oligonucleotide synthesis. Three of the left-side sequence tags (GTA, CTA, TGA) could derived from three longer left-side sequence tags (AGTA, GCTA, ATGA) by N-1 truncation. Only reads with full-length sequence tag sequences were used in this analysis. The three nucleotide-long sequence tags have been redesigned for subsequent experiments to remove any ambiguities that arise from N-1 truncation. The sequence tags were incorporated at the 3` end of the PCR primers. 

\begin{table}[ht]
\caption{\bf{sequence tag and primer sequences}}
\begin{tabular}{rlrl}
L-tag & Name & R-tag & Name  \\\hline
catgcg & 0-v6L & cgcatg & 0-v6R \\
gcagt & 1-v6L & actgc & 1-v6R \\
tagct & 2-v6L & agcta & 2-v6R \\
gactgt & 3-v6L & acagtc & 3-v6R \\
cgtcga & 4-v6L & tcgacg & 4-v6R \\
gtcgc & 5-v6L & gcgac & 5-v6R \\
acgta & 6-v6L & tacgt & 6-v6R \\
cactac & 7-v6L & gtagtg & 7-v6R \\
tgac & 8-v6L & gtca & 8-v6R \\
agta & 9-v6L & tact & 9-v6R \\
atga & 10-v6L & tcat & 10-v6R \\
tgca & 11-v6L & tgca & 11-v6R \\
act & 12-v6L & agt & 12-v6R \\
tcg & 13-v6L & cga & 13-v6R \\
gta & 14-v6L & tac & 14-v6R \\
cta & 15-v6L & tag & 15-v6R \\
tga & 16-v6L & &\\ 
gcta & 17-v6L & &\\
\end{tabular}
\label{barcode}
\end{table}

\subsection*{Extracting sequence reads and sample assignment}

\begin{figure}[ht]
\begin{center}\includegraphics[width=0.7\textwidth]{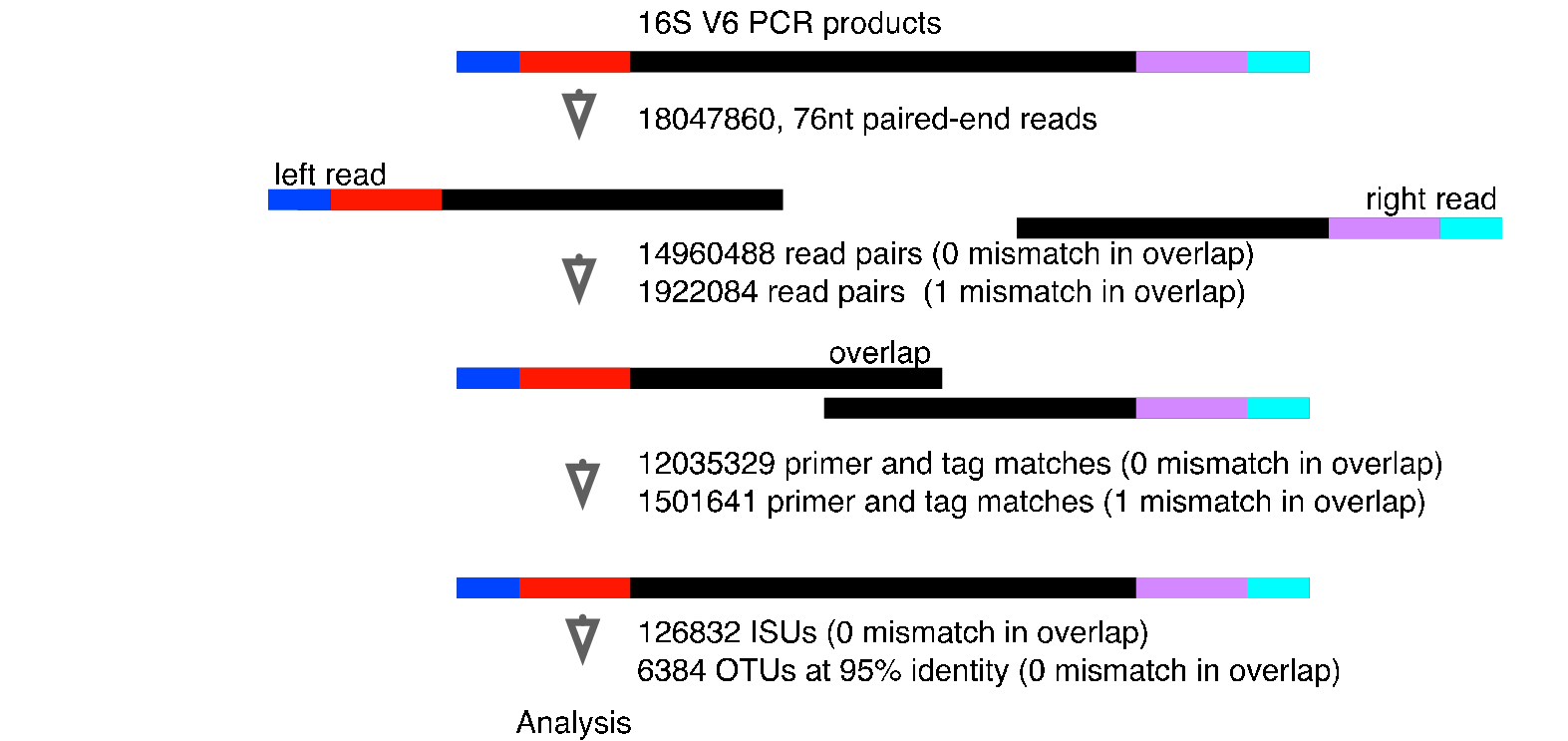}\end{center}
\caption{{\bf Conceptual workflow of the data analysis.} PCR products derived from the eubacterial V6 rRNA region were sequenced on a single paired-end Illumina run. Reads were filtered for quality, overlapped and clustered as outlined in the text. Only reads with 0 mismatches in the overlapping region were used for further analysis.}
\label{workflow}
\end{figure}

As stated by others \cite{Caporaso:2010} the large number of sequences and the short sequence reads present a challenge. The number of and the short length of the reads prevented the application of many commmon pyrosequencing  data analysis pipelines. We therefore developed the data analysis pipeline shown schematically in Figure \ref{workflow}. A full description of each step is given below. All programs to extract the sequence reads and to index them into individual sequence units (ISUs) were developed in-house. A Bash shell script referencing C, Perl and R programs and scripts that are able to recapitulate this analyses on OS X are available from the authors.

We received 18047860 reads that were 76 nt long from each end of the PCR amplified region. Of these, 6236435 and 5491692 reads contained one or more low quality positions in the left and right end reads (defined as having the lowest base quality scores). However, there were only 53598 and 88498 reads that contained one or more `N' character in the sequence calls. 

A custom program was written in C to identify the overlapping segments of the forward and reverse reads. The program first identifies perfect overlaps between the two reads, and then finds reads that overlap if a single mismatch is allowed. The quality score is used to identify the most likely nucleotide in the overlapped segment, and a new fastq formatted file is generated for the combined reads. With this strategy
14960488
reads were obtained that had  a perfect overlap between 10 and 50 nucleotides and an additional 
1922084 
reads had a single nucleotide mismatch in the overlapped region. The extraction of overlapping reads with proper primer sequences and correct sequence tags was performed with a custom Perl program. We found that  
12035329 
sequences contained two valid sequence tags and both primer sequences; allowing up to 3 mismatches per primer. The sequences derived from the perfectly overlapping reads form the basis of the remainder of the analysis. 

Inspection of sequences with incorrect sequence tags showed that the single largest contributor to the difference between the number of reads with proper primer sequences and the number of reads with proper sequence tag sequences was an N-1 truncation of the sequence tag, which presumably arose during the primer synthesis. The next largest class of sequence tag error was complete lack of the left or right end sequence tag. Together, these classes account for slightly more than half of the missing reads. The remaining missing reads are composed of a large number of classes of sequence tag sequences each containing small numbers of errors including additional 5' bases, misincorporated bases or difficult to classify errors that presumably arose during the PCR amplification.

\subsubsection*{Sequence clustering}

Clustering was used to group identical sequences into ISUs, and these ISUs werefurther clustered into operational taxonomic units (OTUs). The variation in an OTU can come from sequence differences between closely related taxa in the underlying population, through errors introduced into the amplified fragment from the PCR amplification, or from DNA sequencing errors. 

ISUs were produced by collecting and collapsing identical sequences located between the primers and collapsing. A custom Perl program was written that associated each ISU  with the number of identical sequences in it, that indexed each read to the appropriate ISU and, later the OTU. The 12035329 reads were collapsed into 126832 ISU sequences, with the most abundant ISU containing 4321348 identical reads.

The occurrence of chimeric sequences was examined using UCHIME, a part of the UCLUST package (REF, http://www.drive5.com/uclust/). Chimeric sequences can arise during PCR \cite{Lahr:2009} or  because of overlapping spots on the solid support when imaged during DNA sequencing \cite{Bentley:2008}. The default settings of UCHIME identified 5211 putative chimeric ISUs, containing 18614 reads. Thus, 5.6\% of the ISUs were putative chimeric sequences, but these composed only 0.15\% of the total reads. There were 21 abundant putative chimeric ISUs that contained $>100$ reads; the most abundant contained contained 1271 reads. 

Each of the abundant putative chimeric ISUs were tested for chimerism with BLAST using the ISU sequence as the query and using the bacterial subset of nucleotide sequences at NCBI as the database. We found only two putative chimeric sequences had sequence  derived from two different species and had a UCHIME chimera score  $>10$, the other 19 putative chimeric ISU sequences matched multiple independent sequences in the dataset with $\ge 98\%$ identity for their entire length. Thus, the occurrence of chimeric sequences was re-evaluated using a chimera score cutoff of 10 and only 497 ISUs containing 1834 total reads (0.015\% of the dataset) were above this threshold. We concluded that chimeric sequences composed a very small subset of the total number of ISUs, probably because the primers amplified across a variable region only. The dataset was used without further regard to chimeric sequences because putative chimeric sequences composed a miniscule fraction of reads. 

The ordered ISU sequences were clustered into OTUs, operational taxonomic units, by UCLUST which clusters each ISU to a seed sequence at a fixed sequence identity threshold using  sequences as seeds in the order they are encountered in the file. We ordered the ISU sequences from the most to the least abundant, under the assumption that read abundance correlated with the abundance of the sequences in the underlying population. Several lines of analysis were used to decide on appropriate clustering values.   

It is expected that the abundance of sequence variants per OTU will decrease according to a power law if the variants are generated stochastically. However, if a variant represents a distinct taxon in the underlying microbial population, the frequency of the variant is expected to reflect the proportion of the bacterial DNA in the sample.

\begin{figure}[ht]
\begin{center}
\includegraphics[width=0.7\textwidth]{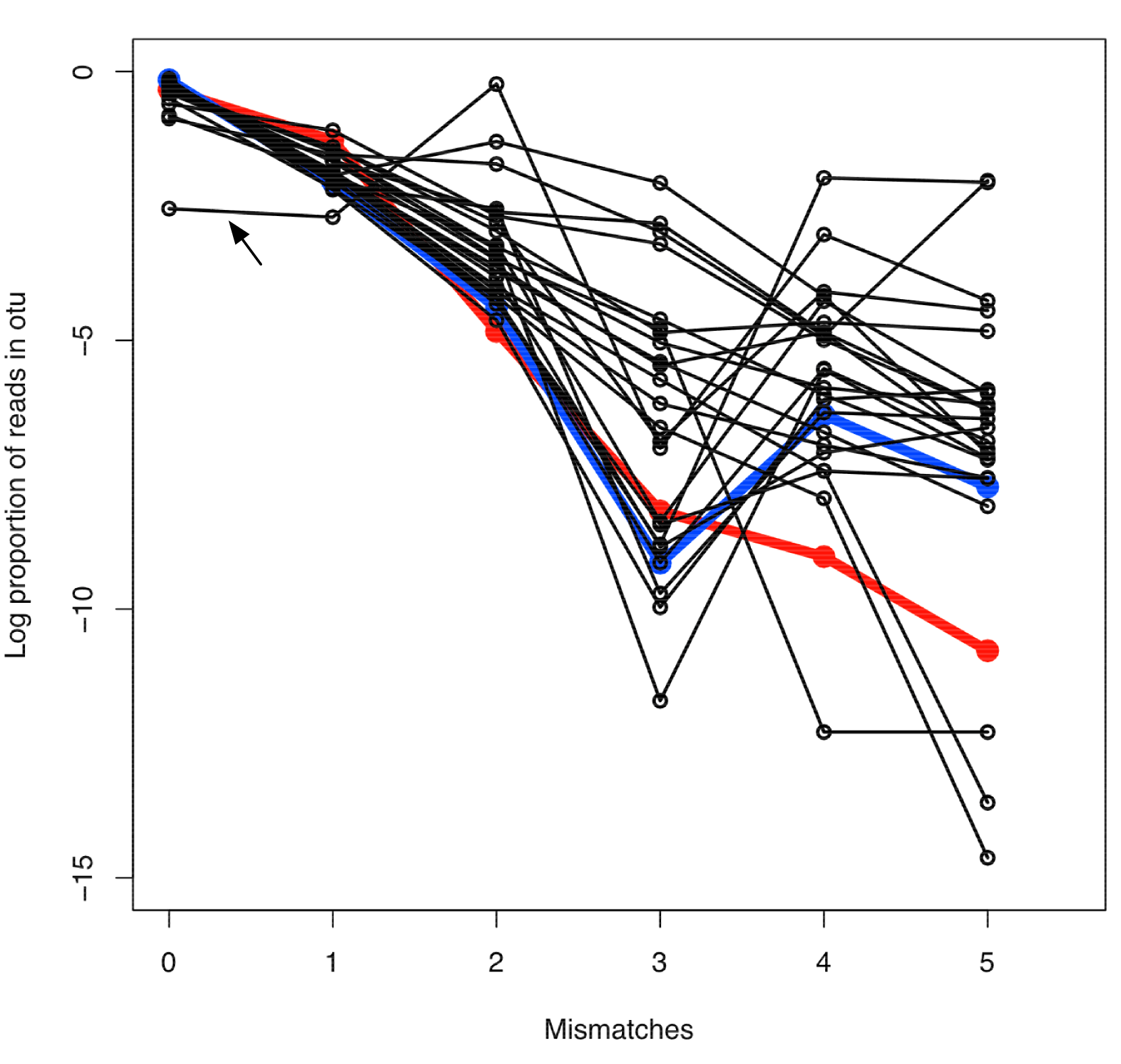}
\end{center}
\caption{{\bf The proportion of reads in the 25 most abundant OTUs clustered at 92\% identity as a function of the number of differences with the seed ISU.} The red line shows the plot for the concatenated primer sequences, and the blue line shows the plot for the OTU containing the most abundant ISU.}
\label{decay2}
\end{figure}

Figure \ref{decay2} shows a plot of the number of reads in an OTU having $n$ mismatches compared to the most frequent read in the OTU at a cluster percentage of 92\%. For an OTU with a length between 72-80 bp this corresponds to $\sim 5$ mismatches with the seed sequence. The red line in Figure \ref{decay2}  shows the plot for the 37 bp concatenated left and right primer sequences, which are expected to have half the per-nucleotide PCR-dependent error rate as the sequence between the primers, because 50\% of the sequence is not derived \emph{de novo} but is contributed by the primer sequence. Because the concatenated sequence is about one-half the length of the sequence between the primers, the overall slope of the primer line should approximate the slope of a single-species OTU that includes errors arising only from the PCR and sequencing. Note that the line for the primer sequence is nearly linear and, in line with our expectations, the number of reads having additional differences with the seed sequence for the OTU is far less abundant than the reads with one fewer difference. Also plotted are the results for the 25 most abundant OTUs, with OTU 0, the most abundant OTU comprising  51\% of the total reads, shown in blue. The line for OTU 0, and several other OTUs closely follow the line for the concatenated primers until 4 or 5 differences with the seed sequence are included. The simplest interpretation is that one or more additional rare taxa having 4 or more mismatches with the seed sequence for OTU 0 are now being included at this level of clustering. The lines for 11 of the 25 OTUs show a similar pattern with a sharp increase at 4 or more mismatches. Only 3 of the OTUs show a continuous decline for all number of mismatches with the seed member of the OTU suggesting that clustering at 92\% identity was including sequences not derived from PCR or sequencing error.

\begin{figure}[!ht]
\begin{center}
\includegraphics[width=0.80\textwidth]{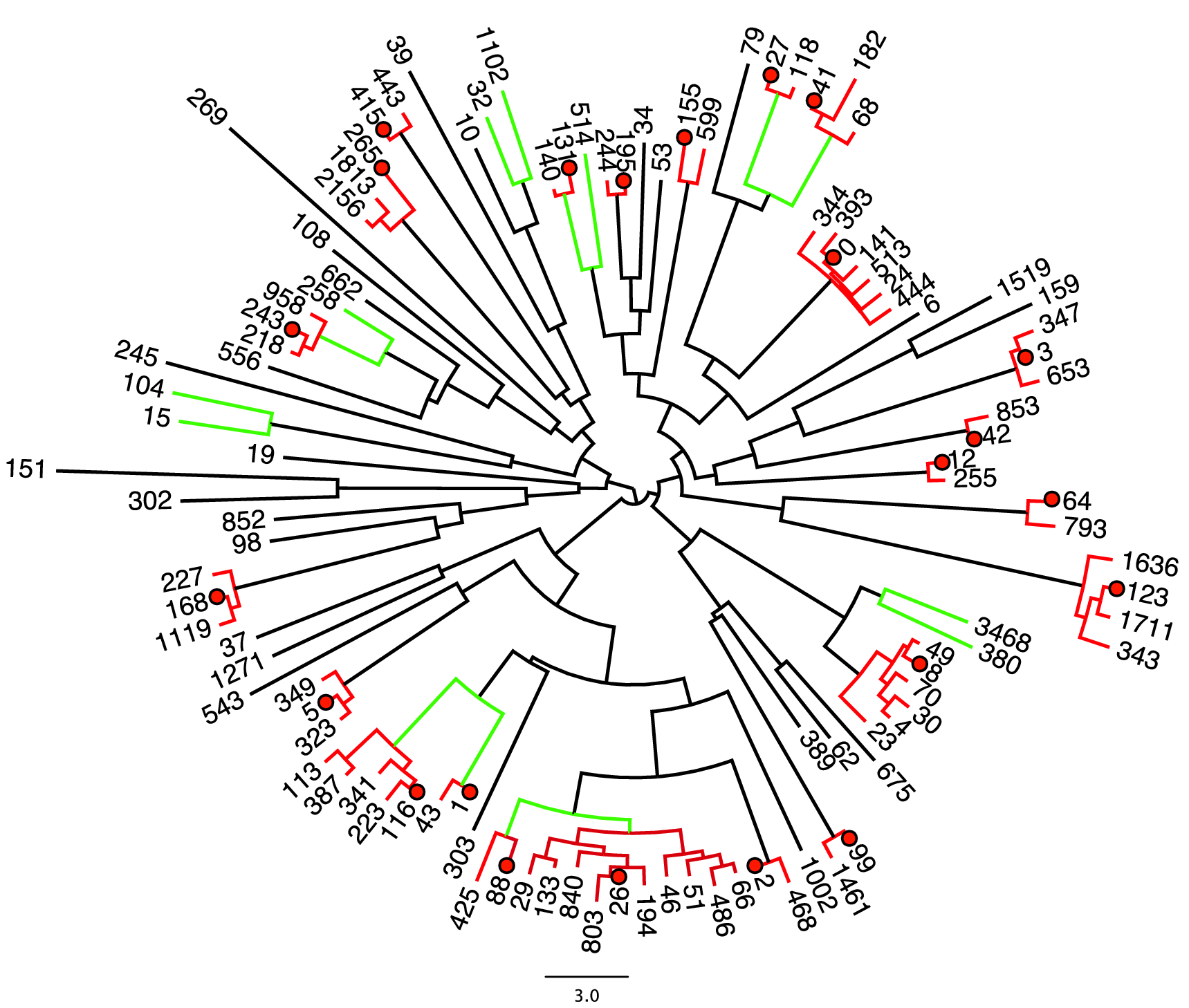}
\end{center}
\caption{{\bf Neighbour-joining tree derived from Levenshtein distance between the 108 most abundant ISU sequences.} ISUs clustered into OTUs at 95\% identity are connected with red branches and ISU sequences clustered at 92\% identity are connected with green branches. The seed sequence for each 95\% identity OTU cluster is identified by a red dot.}
\label{tree}
\end{figure}

We next calculated the Levenshtein distance --- the minimal number of substitutions, insertions or deletions needed to convert one sequence into another --- for all pairs of the 108 ISU sequences that occurred with a frequency of $\ge 1\%$ in any of the 272 samples. Examination of the neighbour-joining tree drawn from these distances showed that there was a continuum of distances between ISU sequences, but that there seemed to be  a natural distance cutoff of three substitutions in this dataset. This is illustrated in Figure \ref{tree} where the branches sharing red nodes connect  ISU sequences that clustered together at 95\% identity, and branches sharing green nodes connect ISU sequences clustered at 92\% identity. Several of these are instructive. The clade at 2 o'clock is anchored around ISU 0. The other ISU sequences in this clade differ from ISU 0 by one or two substitutions, and, as we show below, represent substitutions because of PCR error. All the members of this clade are well-separated from all other clades. The other extreme can be illustrated by the clades at 4 and 6 o'clock. Here, as shown below, the grouping at 95\% identity includes differences derived from PCR errors and from underlying sequence diversity in the microbial sample. However, grouping at 92\% identity (Levenshtein distance of 5) clearly groups outlier clades with the main group. It is standard to assume that clustering at 97\% identity represents species units \cite{Caporaso:2010}. However, taking the two extremes as examples, clustering at greater than 95\% identity would result in splitting clades that contain differences derived only from PCR error (i.e. ISU 0 and associated ISUs) and clustering at less than 95\% identity would group sequences that should be distinct.

Based on these analyses a cluster percentage of 95\% was used for the analysis given below because it allowed up to 3 nucleotide differences with the seed sequence per OTU. At the 95\% clustering threshold,  15 of the OTUs showed ISU mismatch frequency decay characteristics similar to that expected for errors introduced only via PCR or sequencing error; i.e., their abundance profiles decayed at a rate similar to that seen for errors in the primer sequences. This indicates that these 15 OTU sequences may be well differentiated from their neighbours at this level of clustering and may represent distinct sequence species in the underlying population. On the other hand, the most abundant ISU in several OTUs was outnumbered by clustered ISU sequences. In the most extreme cases, OTUs 46, 97 and 119, ISU species with 2 and 3 differences from the seed ISU outnumbered the seed ISU by 2-3 orders of magnitude. An example of this characteristic profile is labeled with an arrow in Figure \ref{decay2}. As shown below, these OTUs represent clusters of errors derived from very abundant organisms in the underlying population. 

\subsection*{Assignment of OTUs to taxonomic groups} The tools used for taxonomic assignment are not designed to work with the short sequences derived from this type of analysis \cite{Cole:2009}. Therefore, similar to others we designed a simple classification scheme based on sequence comparison with BLAST \cite{Camacho:2009, Caporaso:2010} vs. eubacterial sequences (taxid 2), excluding uncultured and environmental samples, in the GenBank database \cite{Benson:2010}. In essence, sequences were identified at the species level if a fully-sequenced or classified type-species sequence matched the OTU with 100\% identity and 100\% coverage \emph{and} no other sequence matched with $>97$\%identity. Sequences that matched with less than 100\% identity were classified at the genus level if another genus matched with a lower percent identity. Sequences with less than 95\% identity were matched to the taxonomic level supported by the groups of reads. With these rules we were able to assign the 63 OTU sequences that were at an abundance of $\ge1$\% in any of the 272 samples unambiguously. As discussed below, three of the OTUs were derived from PCR errors from the \emph{G. vaginalis} strains and were classified accordingly. The classifications of these OTUs, and the supporting evidence for each is shown in Supplementary Table 1.  

\subsection*{Systematic sources of error} 

Recently, Quince et al \cite{Quince:2009} examined the effect of pyrosequencing  errors on the classification of organisms in high-throughput microbiome analyses. They concluded that a large fraction of the `rare microbiota' was composed of pyrosequencing errors and introduced a method to accurately cluster the reads based on their expected errors. Since the Illumina sequencing platform has a substantially lower error rate than does the 454 pyrosequencing platform, and the read length is deterministic rather than random  \cite{Smith:2008} we were thus interested in identifying the sources of error in the $\sim13$ million overlapping reads in our dataset. Most notably the Illumina platform is not susceptible to miscalling the number of nucleotides in homopolymeric regions; this type of base-call error is more pronounced in pyrosequencing reads when sequence coverage is relatively low \cite{Smith:2008}. 

If the major source of error in the data came from DNA sequencing, we would expect that errors should increase as a function of distance from the sequencing primer until the region of overlap and that the errors should be much less frequent in the overlapped region. This hypothesis can be assessed by comparing the $Q$, or quality, score assigned by the Illumina base-calling algorithm in the overlapped 16S sequences with the error frequency per position. 

\begin{figure}[ht]
\begin{center}
\includegraphics[width=0.7\textwidth]{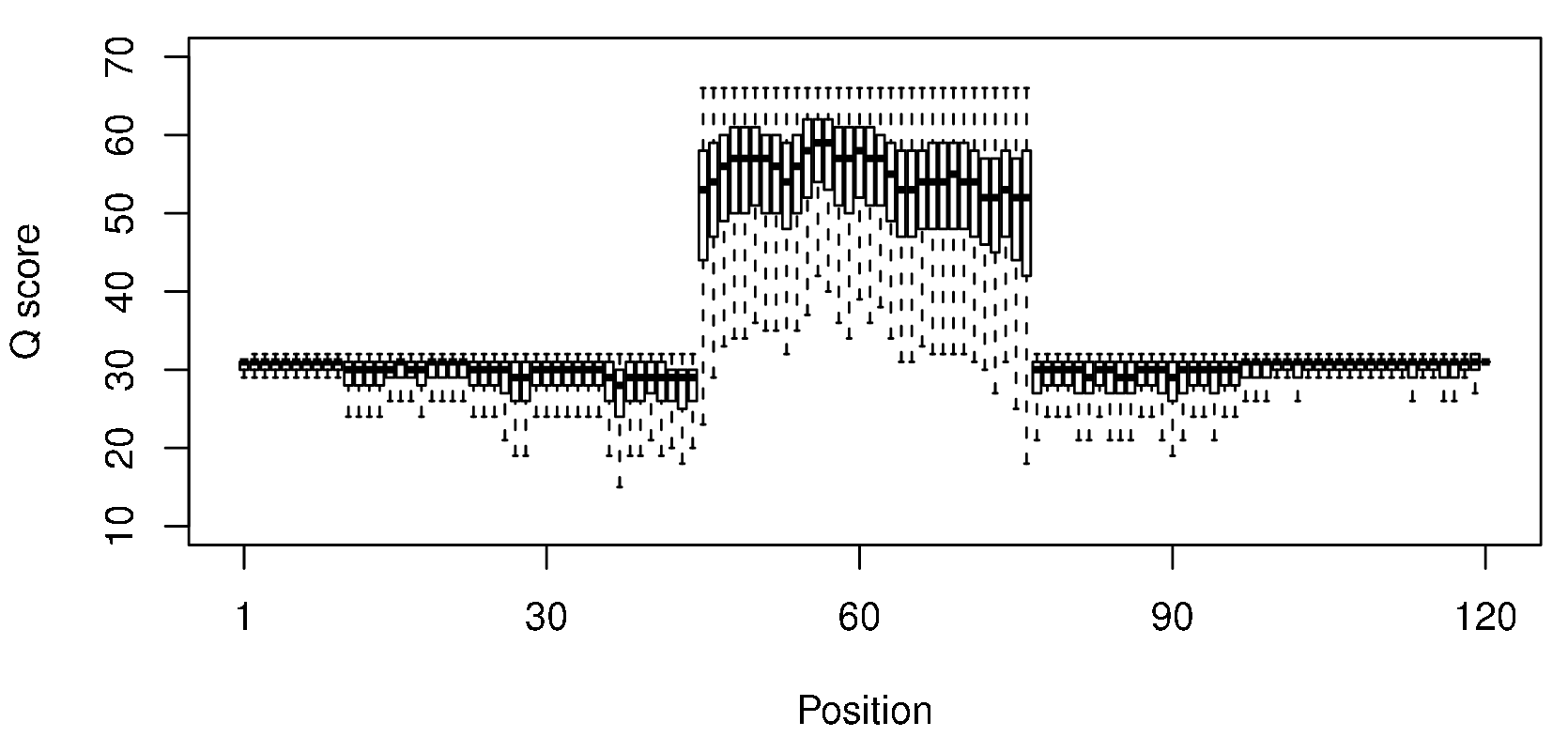}
\end{center}
\caption{{\bf Quality scores for all overlapped 120 bp composite reads.} The $Q$ scores a log-odds score of the likelihood of error in the base call, higher $Q$ scores represent lower likelihoods of error \cite{Cock:2010}. They are expected to decrease with distance from the left or right sequencing primer, and to be highest in the region of perfect overlap because $Q$ scores are additive. }
\label{Q}
\end{figure}

Figure \ref{Q} shows a boxplot of the  $Q$ scores for the reads of length 120 nt, which composed 24\% of the $\sim12000000$ overlapping reads. Similar results were obtained for reads between 113 and 126 nt, which together compose $>99\%$ of the overlapping reads. Two important conclusions can be drawn. First, as expected, the median $Q$ score decreases and the range of scores increase as the distance from the sequencing primer becomes greater. Second, the $Q$ scores, and the variability in these scores for the region of overlap are greater than for the region of single coverage. 

\begin{figure}[ht]
\begin{center}\includegraphics[width=0.7\textwidth]{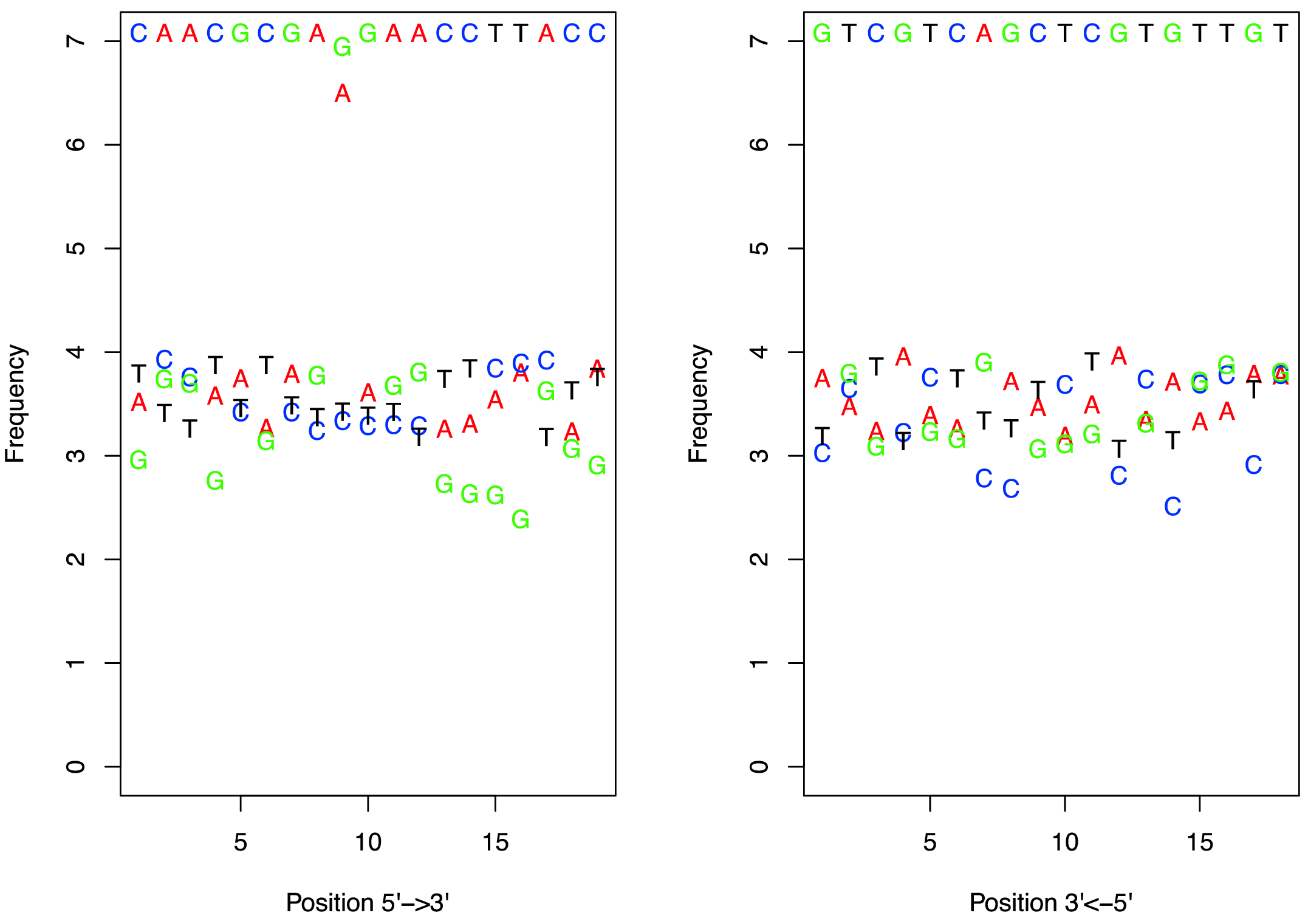}\end{center}
\caption{{\bf The frequency of each nucleotide observed at each position in the left and right primers derived from the Illumina dataset.} There are $> 12$ million sequences, and the difference in frequency between the correct and altered nucleotide is relatively constant. Note that the errors are at the same frequency at each end of the primers.}
\label{primer_mut}
\end{figure}

Initally, the concept of stochastic error contributing to sequence variation was examined by measuring the frequency of occurrence of each nucleotide in the left and right primers. Figure \ref{primer_mut} shows a plot of the number of times  that each nucleotide occurred at each position in the left and right V6 primers. This figure illustrates several points. First, the most frequent variant at each position is usually a transition rather than a transversion, although several positions did not follow this pattern. Secondly, the frequency of the residues differing from the primer sequence are found in a relatively consistent range. Thirdly,  position 9 in the left primer, which was synthesized as a mixture of G and A, shows a strong deviation from the background frequency. Thus the underlying nucleotide frequency in the population of molecules being amplified strongly affects the nucleotide frequency at the polymorphic position. Finally, the variation is constant across the entire length of the primers except for position 9 and is not dependent on the distance from the sequencing primer. These observations support the hypothesis that stochastic errors may contribute significantly to sequence variation in our dataset.

The relationship between the $Q$ scores and the abundance of sequence variants for each OTU was examined by mapping the variants  onto  seed ISUs  as was done for the primer sequences. All ISU sequences in each OTU were used to make a BLAST database for that OTU and the OTU seed sequence was used as the query sequence. An additional 6 nucleotides were added onto both ends of both the OTU sequence and the ISU sequences because of the edge effects in the BLAST algorithm \cite{Altschul:1996}. These  nucleotides were later trimmed for the analysis. The number of sequence variants at each position, weighted by the number of reads that the variant occurred in was tabulated and converted into graphical representations of nucleotide counts at each position in the OTU.  
 
\begin{figure}[t]
\begin{center}\includegraphics[width=0.80\textwidth]{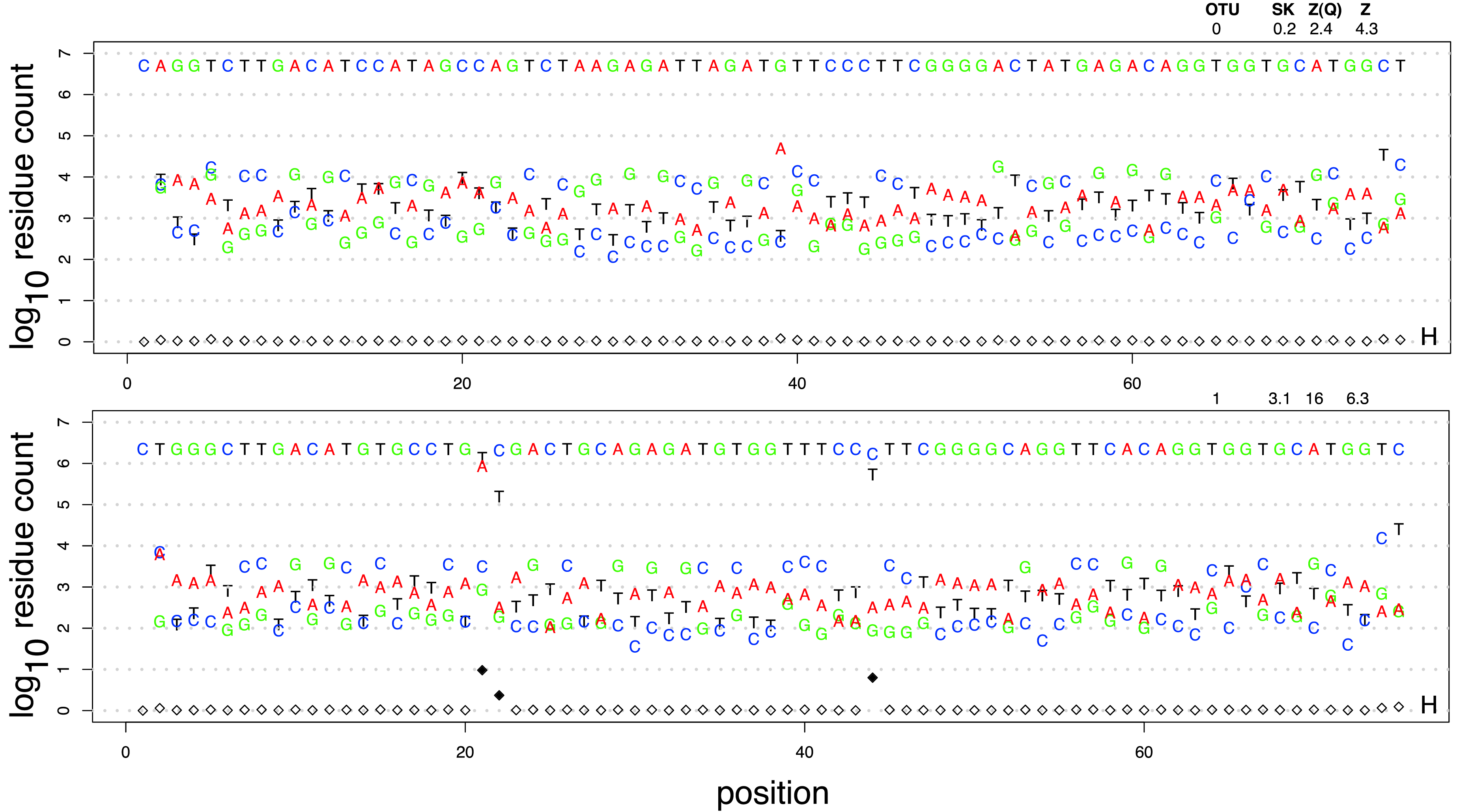}\end{center}
\caption{{\bf The sequence variation in OTU 0 and OTU 1.} The plot shows the number of times that each nucleotide occurred at each position in two example OTUs}
\label{summary}
\end{figure}

\begin{figure}[!ht]
\begin{center}\includegraphics[width=0.80\textwidth]{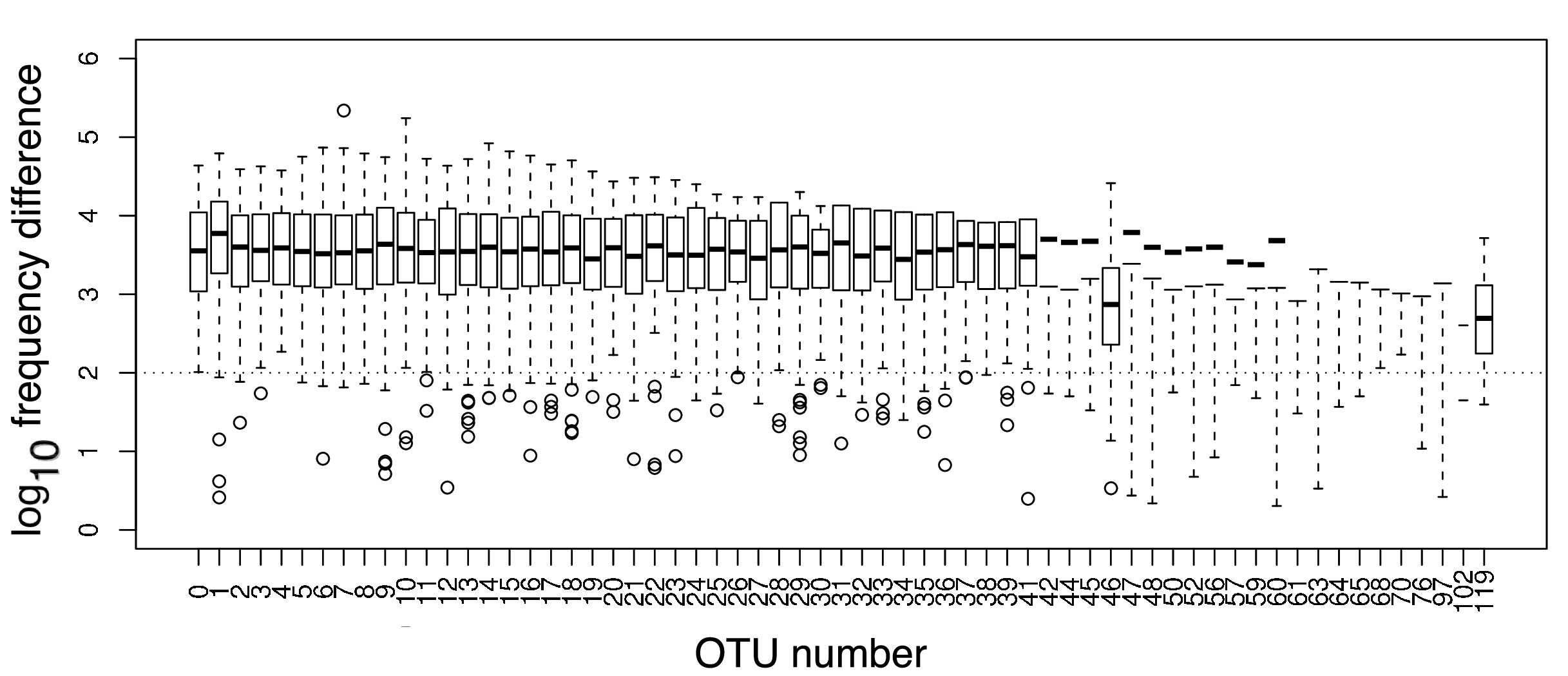}\end{center}
\caption{{\bf Boxplot summaries of the difference between the frequency of the most in common residue at each position and the frequency of each sequence variant.} The OTU numbers are given at the top of the graph.}
\label{residual}
\end{figure}

Two representative examples for the rRNA V6 region are given in Figure \ref{summary}, and a summary of the distributions is given in Figure \ref{residual}.  Figure \ref{summary} shows the number of reads that contain an individual residue at each position plotted in color. The entropy of each position is plotted as open or filled diamonds; higher entropy values correspond to greater variability at the position. Both of these OTUs contain several million reads, and the predominant nucleotide corresponds to the OTU seed sequence. However, there are many variants that were clustered together in these OTUs. 

Figure \ref{residual} shows a summary plot of the distribution of differences in the frequency across each OTU between the most commonly occurring residue and the other 3 residues at each position. The OTUs are arranged approximately from those with the most to the least number of reads.  Several interesting observations can be made from these barplots. First, the frequency differential varies between $10^{-2}$ and $10^{-5}$ for the vast majority of sequence variants from the seed ISU sequence. Second, about half of the OTUs contain  one or more strongly outlying values. These correspond directly to the common variant residues seen in Figure \ref{summary}. Compare, for example, the uniform distribution of variants in the top panel of  Figure \ref{residual} (OTU 0) and the three outlying variants in the bottom panel (OTU 1) with the nucleotide distributions in  Figure \ref{summary} for these OTUs.  Third, the evidence for outlying positions becomes progressively weaker as the the number of sequences in the OTU decreases. 

The data in these two figures can be summarized numerically by examining the distribution of the entropy of the positions in each OTU. Skew in the entropy values is calculated by: $SK = H_{median} - \overline{H} \times 100$. The $SK$ value tells us if the distribution of entropies is strongly skewed by the occurrence of highly variable positions. Values near or greater than 1 indicate a strongly skewed entropy distribution and represent a situation where several to many positions are highly variable.

$Z$ and $Z_Q$ both measure how different the maximum entropy value is to the central tendency of the entropy distribution, and are calculated as follows: $Z = (H_{max} - \overline{H}) / \sigma H$ and  $Z_Q = (H_{max} - H_{median} )/ (H_{95^{th}percentile} - H_{median})$. Thus $Z$ represents the number of standard deviations that the maximum entropy value is from the mean, and $Z_Q$ is the number of 95 percentile deviations of the maximum entropy value  from the median.  Both values are required since $Z$ is not informative if a distribution has a large variance. $Z_Q$ has extreme values in the instances of a skewed distribution with small number of extreme values. Inspection of the plots suggests that values of$SK>1$, $Z> 6$ or $Z_Q>6$ represent situations where the nucleotide distribution for a suggests a mixed population of reads. Conversely, OTUs where all 3 values are less than these cutoffs strongly suggest that the  variability seen in the OTU arose from stochastic errors inherent in the experimental protocol.


The information in Figure \ref{summary} and in \ref{residual} and the associated entropy information, allow us to classify the OTUs into groups that contain a homogeneous population of reads that differ from each other only because of variations introduced during the PCR step (eg. OTU 0) and OTUs that contain sequence variants derived from the underlying population (eg. OTU 1). 

As an example, the top panel in Figure \ref{summary} corresponds to OTU 0, and the seed sequence in this OTU is identical to the V6 region of \emph{Lactobacillus iners} in both the RDP and NCBI nucleotide databases. The bottom panel corresponds to OTU 1, and the seed sequence is identical to one annotated as \emph{Gardnerella vaginalis 409-05}. The second most common sequence is identical to one annotated as \emph{G. vaginalis NML060420}, and the third and fourth most common sequences are identical to sequences annotated as uncultured \emph{G. vaginalis} sequences. All four of these sequences differ from each other by a single diagnostic nucleotide, and the variant counts match the counts of the 4 major ISUs. These 4 ISUs make up the 88.9\% of the reads in OTU 1. Based on the analysis of these two OTUs and the similar analyses of the remaining OTUs, we conclude that OTUs that exhibit the pattern of variation seen in OTU 0 represent distinct sequence entities in the underlying dataset and that those exhibiting a pattern of variation similar to OTU 1 represent the grouping of sequence entities in the underlying dataset based on sequence similarity. In the case of OTU 0, no sequence in the RDP database \cite{Cole:2009} could  be clustered with it without including at least 5 nucleotide substitutions, leading us to conclude that OTU 0 represents a distinct taxonomic group at the sequence level. In the case of OTU 1 there are several sequences, all annotated as different strains of the same species that are grouped together, and like OTU 0, all are well-separated from other V6 sequences. Thus, we conclude that OTU 1 is a cluster of distinct \emph{G. vaginalis} strains. 

OTUs  46, 97 and 119 in the dataset, had distinct distributions when plotted as in Figures \ref{summary} and \ref{residual}. The nucleotide frequency difference between the seed sequence and the nucleotide variants in these three OTUs was much smaller than in the other 61 OTUs. Inspection of the sequences making up these OTUs showed that they were most similar to one or more of the \emph{G. vaginalis} strains. We propose that these OTUs are composed of ISU sequences derived solely from PCR errors that failed to cluster with the seed sequence in OTU 1. We are currently working on a clustering procedure that explicitly accounts both for edit distance and read abundance to more accurately cluster sequences derived by very high throughput sequencing.  

\begin{figure}[ht]
\begin{center}\includegraphics[width=0.7\textwidth]{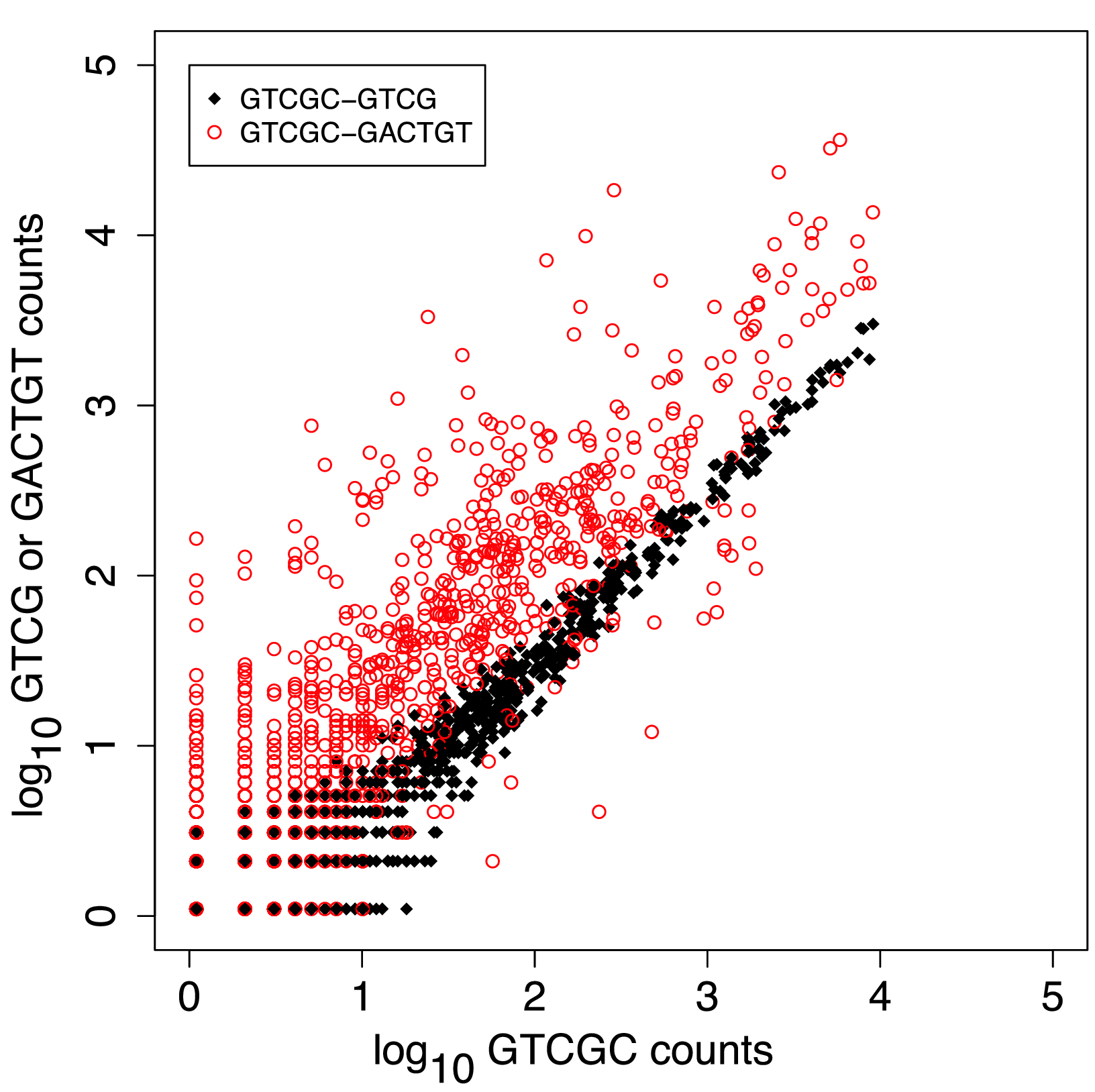}\end{center}
\caption{{\bf Plot of the reproducibility between and within samples.}The black-filled circles plot within-sample variation, and the red circles plot the between-sample variation for the GTCGC tag. The count of sequences composing OTUs clustered at 95\% identity for samples containing the GTCGC tag and the GTCG N-1 tag are in black. This shows the technical replication of the data when amplified from the same sample in the same tube. The open red circles plot the correspondence for between-sample OTU counts.}
\label{reproducibility}
\end{figure}

\subsection*{Organism diversity and data reproducibility}

We found that one right-end tag, GCGAG, was composed of a mixture with the ratio  69.5/30.5 of the full-length and the unique N-1 truncation-derived GCGA tag. This oligonucleotide synthesis error was exploited to determine the effect of the number of reads on within-sample variability; in essence the N-1 truncated tag allowed an examination of  the technical replication of the experiment. The GCGAC tag was used in 17 samples. The black-filled circles in Figure \ref{reproducibility} show the number of reads from the full length GCGAC tag compared to the truncated GCGA tag in these samples. The red open circles in Figure \ref{reproducibility} show an example of the read replication observed from independent samples. The replication of the read numbers in the full length and N-1 samples is extremely high for reads occurring at least 30 times in the full-length tag set, and at least 10 reads in the N-1 tag set.  As expected the read replication for independent samples is much poorer. The correlation coefficients for the 17 full-length and N-1 samples ranged from 0.97 to 0.99 when the N-1 sample contained at least 10 reads. Thus, we conclude that the number of reads in a sample is reproducible, if at least 10 reads are observed. Similar conclusions about the minimum read abundance have been drawn from RNA-seq experiments \cite{Mortazavi:2008}.  

\begin{figure}[!ht]
\begin{center}\includegraphics[width=0.80\textwidth]{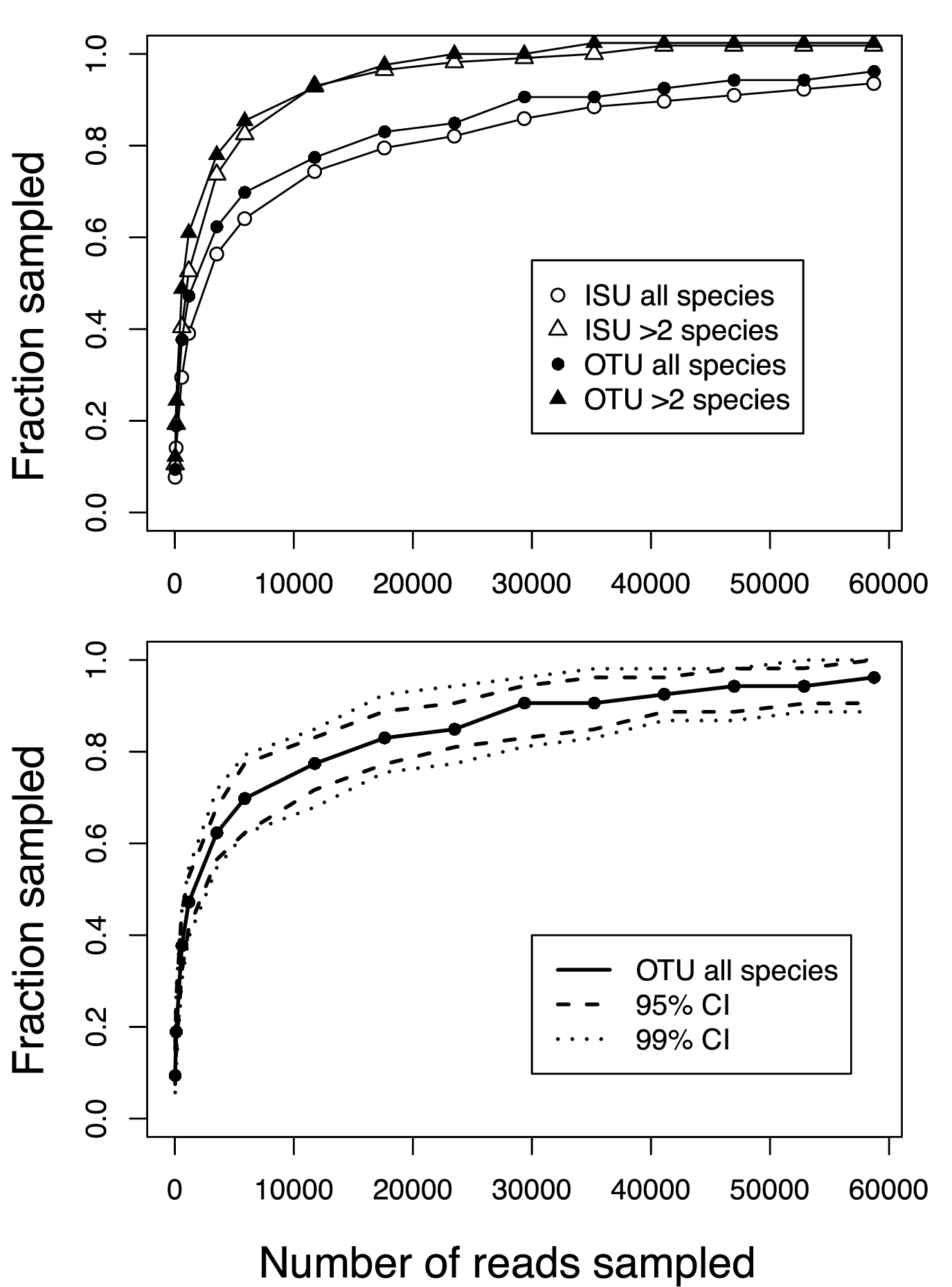}\end{center}
\caption{{\bf An example rarefaction curve.} The top panel shows  rarefaction curves generated for sample 1 by resampling with replacement either all OTUs or ISUs,  or OTUs and ISUs where at least 3 reads were observed. The bottom panel shows the rarefaction curve and the 95\% and 99\% confidence interval for all OTUs in sample 1. 
}
\label{rarefaction}
\end{figure}

\subsubsection*{Rarefaction curves}

A second way to examine reproducibility is to generate rarefaction curves  where the number of species sampled per unit of effort is estimated by resampling the dataset \cite{Hurlbert:1971,Smith:1984}. Rarefaction curves for the dataset from each sample were generated by performing 10000 random samples with replacement \cite{Efron:1981} on the complete set of OTUs or ISUs or by including only those OTUs and ISUs that occurred in a sample more than twice. 
The values for resampling without replacement will approach the observed value (i.e. will saturate) only if the sample is of  sufficient size to encapsulate all possible diversity \cite{Efron:1981}. Thus, if the values do approach saturation when  resampling with replacement, we can be confident that we have sampled most, if not all, of the available sequence species \cite{Efron:1981,Colwell:2009}. 

Figure \ref{rarefaction} shows rarefaction curves generated for ISUs and OTUs in sample 1  using different protocols for a representative sample in our dataset; it is worth pointing out that this rarefaction curve is one of the few curves that does not reach saturation. The white-filled symbols show curves generated for unclustered ISUs in this sample, and the black-filled symbols are for OTUs generated at 95\% sequence clustering. Here, the effect of removing rare sequence species is clear. The curve saturates when sampling only 50\% of the reads if either rare ISU or rare OTU sequences are removed, but does not saturate for either the ISUs or OTUs even with the full set of reads. Inspection of the full set of rarefaction curves shows that this failure to reach the limit is commonly observed when the sample is dominated by one or a few species, which is the case in many of the microbiota samples in our complete dataset. Samples containing a broader range of species show rarefaction curves that generally reach the limit  near 20000 reads, suggesting that this is an appropriate number of reads to sample the microbiome in the vaginal environment. 

\begin{figure}[!ht]
\begin{center}\includegraphics[width=0.7\textwidth]{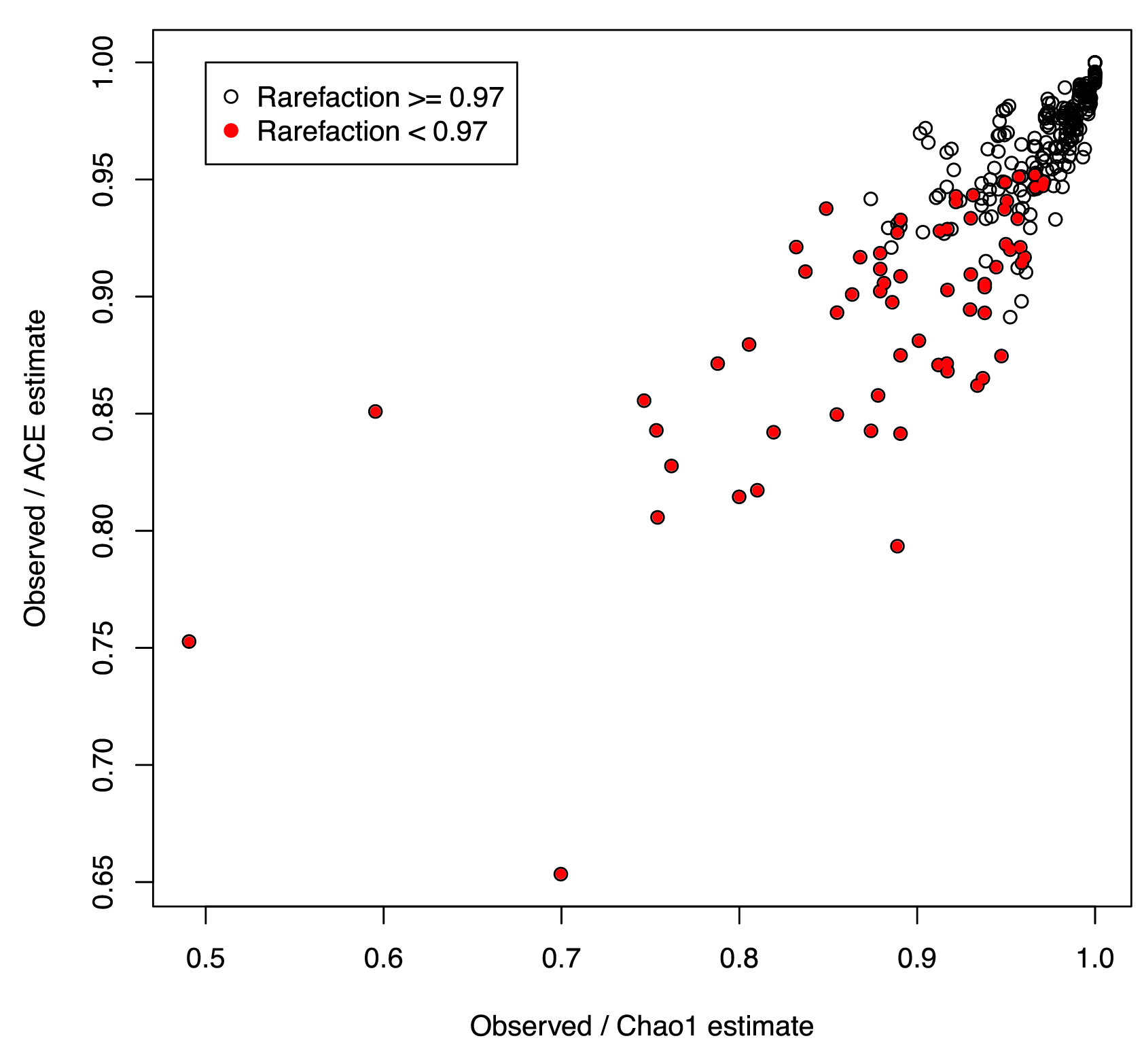}\end{center}
\caption{{\bf Correspondence between Chao1, ACE and rarefaction curves for the 272 samples.} The X and Y axes show the fraction of species that were found in each sample for the two estimates. Red-filled circles highlight those samples where the limit rarefaction value was less than 0.97. }
\label{chao}
\end{figure}

\begin{figure}[!ht]
\begin{center}\includegraphics[width=0.7\textwidth]{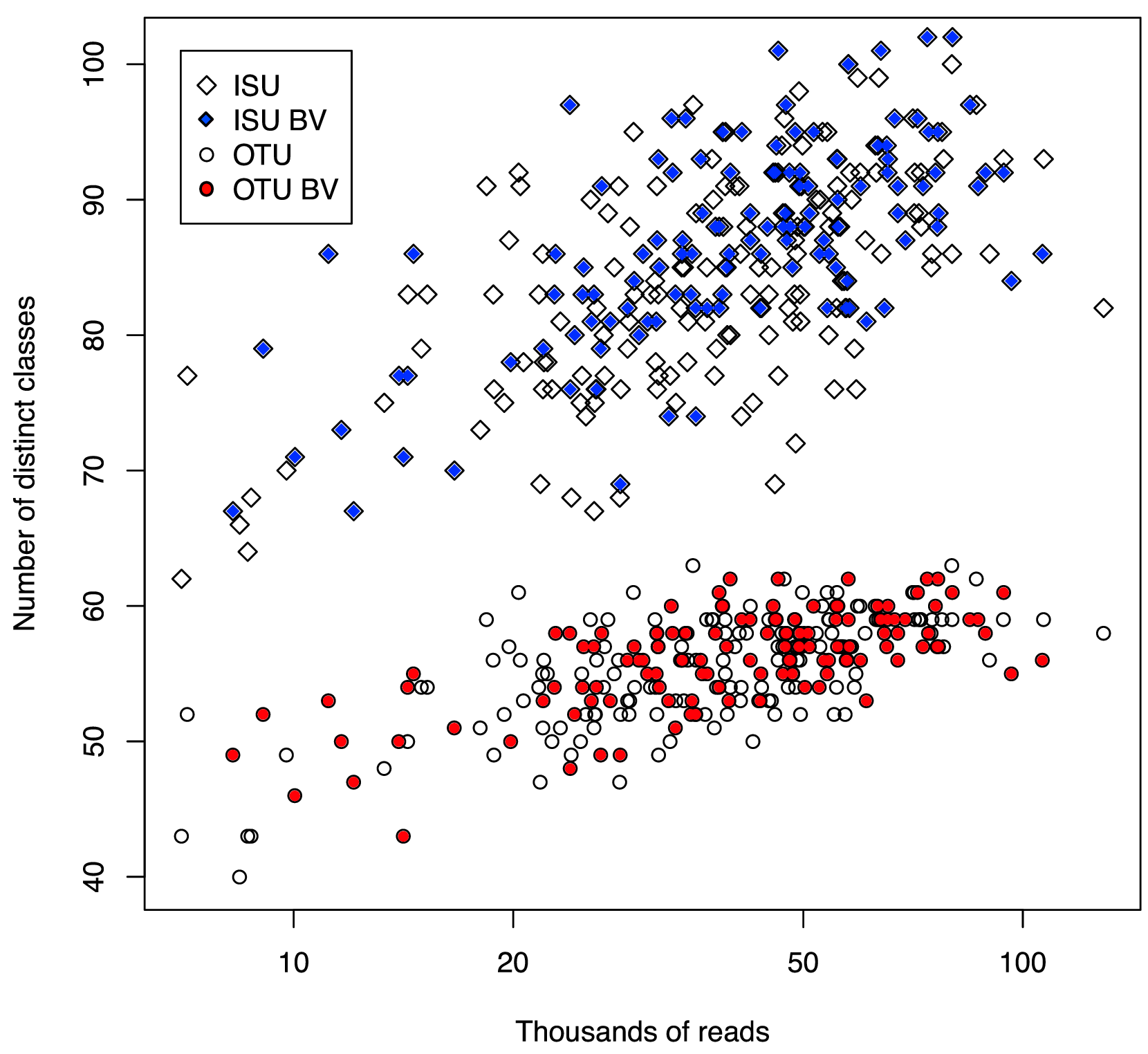}\end{center}
\caption{{\bf Plot of the number of distinct ISU or OTU classes in each sample as a function of the number of reads.} The number of ISU classes increases with the number of reads, but the number of OTU classes becomes constant above 20000 - 30000 reads. }
\label{richness}
\end{figure}

\subsubsection*{Estimating species richness}

Another method of examining species richness is to use the Chao1 or ACE methods to estimate the number of  unseen species in the sample \cite{Chao:1984, Chao:1992}. We used both methods to determine the number of species expected in each of the 272 samples with the VEGAN package for biodiversity analysis \cite{Oksanen:2010}. There were 37 and 31 of 272 samples where the Chao1 and ACE estimates indicated that we observed $<90\%$ of the real species. The correspondence between the Chao1 and ACE measures is plotted in  Figure \ref{chao} and it is clear by these two measures that the vast majority of samples were expected to contain the majority of the available species. Included in  this plot is the fraction of species found when the rarefaction analysis was performed with the number of reads in the sample. Rarefaction with a saturating number of reads again showed that the 206 of 272 samples identified all or almost all of the available species.  

\subsubsection*{Diversity vs. number of reads}

Finally, species richness can be examined as a function of the number of reads across all 272 samples. This is plotted in Figure

\ref{richness} for ISU and OTU sequences. In this case the white-filled symbols represent populations derived from samples classified as `normal', and are expected to be dominated by one or a few species, and the red or blue-filled symbols represent populations classified as bacterial vaginosis (BV), where there is expected to be a more even distribution of species \cite{Srinivasan:2008, Shi:2009}. 
There are strongly diminishing returns when more than 20000-25000  reads are obtained regardless of the diversity of the population; sampling more than 50000 reads was sufficient to sample all the available OTU diversity in the samples. Interestingly, the number of distinct ISU sequences increases linearly with the number of reads, providing further evidence that increasing the number of reads increases the background number of ISUs that contain PCR-derived errors.   Taken together with the rarefaction, Chao1, ACE data, we conclude that the number of reads obtained by this Illumina sequencing is adequate to sample nearly saturating numbers of species in this environment.  

\begin{figure}[ht]
\begin{center}\includegraphics[width=0.7\textwidth]{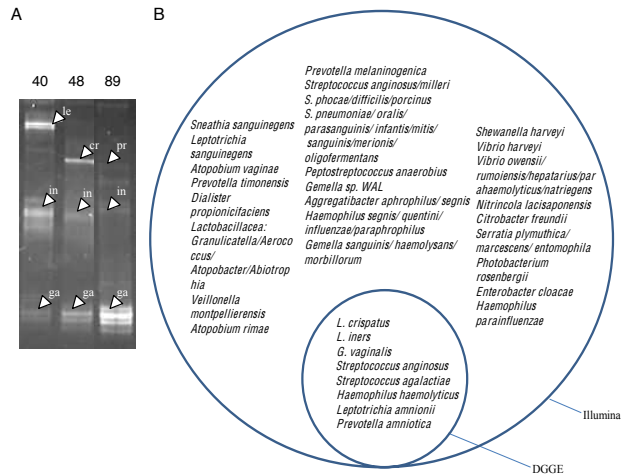}\end{center}
\caption{{\bf DGGE analysis of selected samples.} Panel A shows representative PCR amplicons from 3 of 20 clinical samples (Subjects 40, 48 and 89) were electrophoresed on a denaturing gradient gel. Bands were excised, sequenced and identified as in the Materials and Methods. Bands are labeled as follows: le=\emph{Leptotrichia amnionii}; in=\emph{Lactobacillus iners}; ga=\emph{Gardnerella vaginalis}; cr=\emph{Lactobacillus crispatus}; pr=\emph{Prevotella amniotica}. Panel B shows a Venn diagram of the organisms identified by Illumina sequencing of the V6 rRNA region and by sequencing DGGE bands amplified from the V3 rRNA region.}
\label{venn}
\end{figure}

\subsubsection*{Comparison with DGGE} Results from Illumina sequencing were compared to those from dideoxy chain termination sequencing of bands isolated from following denaturing gradient gel electrophoresis (DGGE) analysis of amplified PCR products; a method traditionally used for the separation of bacterial species. A total of 20 samples were selected that were expected to have a diverse population of organisms by extrapolation from the 272 samples sequenced by Illumina. DNA fragments from the bands were sequenced and each OTU sequence and each sequence from the DGGE bands were assigned to taxonomic groups by BLAST using the GenBank nucleotide database as described above. OTU sequences were assigned to species if they matched 100\% of their length at 100\% identity, and to genus or other groups as outlined in Supplementary Table 1. DGGE was found to detect only those bacterial species of greatest abundance in the sample, with a minimal Illumina read abundance of 11\%. In two cases, one shown in Figure \ref{venn}A-lane 89, a distict band was excised and sequenced that had an Illumina abundance of between 2-3\%. Figure \ref{venn}B shows that a total of 8 organisms were detected through DGGE analysis, compared to 59 organisms detected through Illumina analysis in the same 20 samples, and that the organisms identified by DGGE analysis were a strict subset of those identified by Illumina sequencing.

\section*{Discussion}
We present and characterize a low-cost, high throughput method for microbiome profiling. The method uses combinatorial sequence tags attached to the 3` end of PCR primers that amplify the rRNA V6 region, but may be easily adapted for use in other bacterial and eukaryotic sequences.  Illumina paired-end sequencing of the amplimers generates millions of overlapping reads. The combinatorial sequence tags allows the investigator to examine hundreds of samples with far fewer primers than is required for single-end bar-code sequencing. We propose that this method will be useful for the deep sequencing of any short sequence that is informative; these include the V3 and V5 regions of the bacterial 16S rRNA genes and the eukaryotic V9 region that is gaining popularity for sampling protist diversity. The use of the V3 and V5 regions is currently being attempted by our group. 

A recent report used a small number of sequence tags on one end of one primer to examine microbial diversity using Illumina sequencing \cite{Caporaso:2010}. However, this method required three Illumina sequencing to fully identify the sequence tags, and the tags were much longer. In our study, we used a simple set of rules to choose sequence tags that balanced the nucleotide composition in the first 4 positions of the reads, that maximizee the stagger in the primer sequences when attached to the solid surface and that minimized the possibility of primer-dimer formation. Using these simple principles and avoiding the N-1 generation of non-unique sequences, short sequence tags should be easily derived that are suitable for primers specific to any small region of interest. Sequence tags can be chosen automatically using the barcrawl program \cite{Frank:2009} or can be chosen by hand. 

We observed very few chimeric sequences in our dataset. There are several reasons. First, we used a relatively small amount of input DNA and used a small number of PCR cycles for amplification \cite{Lahr:2009}. Secondly, many chimeric sequences may have been removed because of the strict requirement for proper sequence tag and primer sequences on the left and right ends, and because of a requirement for long overlapping segments of a defined length. In this case, the deterministic read lengths of the Illumina protocol combined with our narrow window for overlapping segments would have resulted in many chimeric sequences being filtered out. Indeed, inspection of a fraction of the read pairs that failed to overlap, or that failed to pass the sequence tag and primer requirements showed that many of these were chimeric or deleted at one or both ends (data not shown).  Thus, while the Illumina sequencing protocol is limited to short segments these can be combined into longer segments using the paired-end approach as long as there is a significantly overlapping segment.

The utility of the method is further demonstrated by the near-saturating number of ISU and OTU sequences obtained from a large number of clinical samples. We used several lines of evidence to show that 20000 reads are sufficient to capture all or virtually all of the sequence diversity in the vaginal microbiome, and that obtaining over 50000 reads results in no new sequence species. Thus, assuming a requirement for 50000 reads, up to 200 samples can be combined into a single Illumina lane, while up to 500 samples are possible if only 20000 reads are required. This is much greater depth at a much lower cost than is possible with current pyrosequencing technology. Strikingly, we observed that none of our samples contained the full range of species in the microbiome as a whole, and that we found fewer species than in a recent report that used pyrosequencing in the same niche  \cite{Ravel:2010}, despite averaging 20-fold greater sequence coverage. We suggest that the higher fidelity Illumina sequencing may have resulted in fewer taxa because of a lower error rate contributing to fewer `rare microbiome' taxa.

Finally, we showed that the spectrum of errors could be examined for each OTU to help determine if the OTU was derived from a single underlying sequence in the sample population.    The large number of reads presented a challenge for sequence-based clustering because sequencing millions of reads ensured that much of the read variation  was derived from PCR-amplification. We show that sequence clustering of the large number of reads derived from Illumina sequencing would be more accurate if it took both the sequence variation and the underlying error rates into account. We are currently working on developing methods to cluster that use both sequence similarity and read abundance.

\section*{Materials and Methods}
\subsection*{Ethics Statement} The medical ethical review committee of Erasmus University Medical Centre, The Netherlands, and the medical research coordinating committee of the National Institute for Medical Research, Tanzania, approved the study design and protocol. Subjects were informed of the purpose of the study and gave their signed informed consent before participation. The study was registered at clinical trials.gov NCT00536848.

\subsubsection*{Sample preparation and amplification} DNA was prepared from clincal swabs as part of a clinical study \cite{Hummelen:2010}. Amplification was initiated with a $61^{\circ}$ annealing temperature that dropped to $51^{\circ}$ in $1^{\circ}$ increments followed by 15 cycles of: denaturation $94^{\circ}$, annealing $51^{\circ}$, extension $72^{\circ}$ all for 45 seconds with a final elongation for 2 minutes. A constant volume aliquot of each amplification was run on a 1.4\% agarose gel for to determine the approximate amount of product. The amount of product was scored on a 4 point scale and, based on this scale, between 2 and 40 $\mu l$ of the PCR products were mixed together to give the final sample sent for Illumina sequencing at The Next-Generation Sequencing Facility in The Centre for Applied Genomics at the  Hospital for Sick Children in Toronto. The library was prepared without further size selection. 

\subsubsection*{Denaturing gradient gel electrophoresis analysis}
Clinical samples were amplified using eubacterial primers flanking the V3 region of the 16S rRNA gene: HDA-1 (5Õ-ACTCCTACGGGAGGCAGCAGT-3Õ) at position 339-357 (with a GC clamp located at the 5Õ end), and HDA-2 (5Õ-GTATTACCGCGGCTGCTGGCA-3Õ) at position 518-539, with an annealing temperature of $56^{\circ}$C. PCR reactions were carried out in 50 $\mu l$ reactions for 30 cycles using the profile: $94^{\circ}$C, a gradient of annealing temperatures $71-51^{\circ}$C at 45sec each, elongation $72^{\circ}$C all for 45sec.

Preparation of the 8\% polyacrylamide denaturing gradient and gel electrophoresis was done according to the manufacturerÕs instructions for the D-Code Universal Detection System (Bio-Rad) with a 30-50\% gradient of urea and formamide. The gel was run in  Tris-acetate buffer and pre-heated to $59^{\circ}$C. The  gel was run at 130V for 2 hours or until the xylene cyanol dye front reached the lower end of the gel. DNA was visualized by UV irradiation following stain with ethidium bromide. Bands were excised and reamplified, using the same primers and profile but without the GC clamp. This second PCR product was purified and sequenced with the HDA forward primer via dideoxy chain termination. Analysis of results was carried out using the GenBank nucleotide database and BLAST algorithm \cite{Altschul:1997}.

\section*{Acknowledgments}

We thank Dr. Sergio Pereira (The Center for Applied Genomics, University of Toronto), and Layla Katiree from Illumina for insights into the inner workings of the Illumina platform. The design and analysis was done to analyze the microbiome of a large cohort of HIV+ women in Tanzania. Partial funding of this project from the National Sciences and Engineering Research Council of Canada (NSERC) to Gregor Reid and Greg Gloor is acknowledged. AF is supported an NSERC post-doctoral fellowship and RD is supported by an NSERC CGS-D scholarship. The funders had no role in study design, data collection and analysis, decision to publish, or preparation of the manuscript.

\bibliography{bibdesk_refs}

\end{document}